\documentclass[galaxies,article,submit,moreauthors,pdftex,10pt,a4paper]{mdpi} 
\firstpage{1} 
\articlenumber{x}
\doinum{10.3390/------}
\pubvolume{xx}
\pubyear{2016}
\copyrightyear{2016}
\externaleditor{Academic Editor: name}
\history{Received: date; Accepted: date; Published: date}
\newcommand{\enzo}{\it{\small ENZO}}

\Title{The challenge of detecting intracluster filaments with Faraday Rotation}

\Author{Nicola Locatelli $^{1,2}$, Franco Vazza $^{1,2,3}$,P. Dom\'inguez-Fern\'andez$^{3}$ }
\AuthorNames{Nicola Locatelli, Franco Vazza \& Paola Dom\'inguez-Fern\'andez }

\address{%
$^{1}$ \quad Dipartimento di Fisica e Astronomia, Universit\'{a} di Bologna, Via Gobetti 93/2, 40129 Bologna, Italy\\
$^{2}$ \quad INAF-Istitituto di Radio Astronomia, via Gobetti 101, 40129 Bologna, Italy\\
$^{3}$ \quad Hamburger Sternwarte, Universit\"{a}t Hamburg, Gojenbergsweg 112, 41029 Hamburg, Germany}
\corres{Correspondence: nicola.locatelli2@unibo.it}

\abstract{The detection of filaments in the cosmic web will be crucial to distinguish between the possible magnetogenesis scenarios and future large polarization surveys will be able to shed light on their magnetization level. In this work, we use numerical simulations of galaxy clusters to investigate their possible detection. We compute the Faraday Rotation signal in intracluster filaments and compare it to its surrounding environment.
We find that the expected big improvement in sensitivity with the SKA-MID will in principle allow the detection of a large fraction of filaments surrounding galaxy clusters. However, the contamination of the intrinsic Faraday Rotation of background polarized sources will represent a big limiter to the number of objects that can be significantly detected. We discuss possible strategies to minimize this effect and increase the chances of detection of the cosmic web with the large statistics expected from future surveys.}

\keyword{galaxy: clusters, general -- techniques: polarimetric -- intergalactic medium -- large-scale structure of Universe}

\begin{document}

\section{Introduction}
Magnetic fields in the Universe are observed to permeate a very wide range of spatial scales, from planetary ($\sim 10^6$m) to galactic ($\sim$kpc), up to galaxy cluster scales ($\sim$Mpc). On the largest scales however, their value is constrained only in the regions where the plasma conditions of density and temperature allow their observation, i.e. in the Intra Cluster Medium (ICM, e.g. \citet{fe12} and references therein for a review).
The amplification of magnetic fields in cosmic structures might have proceeded in a bottom-up small-scale turbulent dynamo \citep[e.g.][]{ry08}.  Under such conditions, the dynamo should have erased most traces of the initial magnetization seeds,  bringing the magnetic energy density close to equipartition with the plasma kinetic energy in galaxy clusters. This scenario well explains the observed magnetic fields in galaxy clusters,  i.e. for densities and temperatures of $n \geq 10^{-4}$ [cm${}^{-3}$] and $T \sim 1-10$~keV, respectively,  as supported by numerical simulations \citep[e.g.][]{do99,va18mhd,review_dynamo}.

On the other hand, the magnetization in the intergalactic medium (IGM) outside  the virial radius of galaxy clusters ($n<10^{-4}$ cm${}^{-3}$ and temperature $\sim 10^{5}-10^{7}$ K), is instead still largely unconstrained, despite the fact that this gas phase should contain a large fraction ($\sim 50-60\%$) of the baryonic mass of the Universe.
Measuring the magnetic field intensity and morphology in the IGM is crucial to constrain the primordial origin of extragalactic fields, out of which present day magnetic structures might have evolved  \citep[e.g.][]{wi11}.
Unluckily, direct observations of magnetic fields in the IGM are made  challenging by the very high sensitivity required for the imaging at most wavelengths. 

~Among the techniques nowadays available  to study extragalactic  magnetic fields there are several promising algorithms which apply to the rotation measure (RM) of linearly polarized signals emitted by radio galaxies. In particular, RM Synthesis \citep[][]{2005A&A...441.1217B} is a powerful tool to extract valuable information from both magnetized layers and emitting sources along the line-of-sight (LOS). Attempts to constrain  extragalactic magnetic fields  on statistical bases have set upper limits at the level of $0.3-7 \,\mu$G \citep[][]{2006ApJ...637...19X,2015A&A...575A.118O}{\footnote{Upper limits in the same range have been reported by statistical studies of cross-correlation between radio surveys and galaxy catalogs by \citet{vern17} and \citet{brown17}.}}. Very recently, the possible detection of the RM contribution by filaments overlapping the polarised emission by a giant radio galaxy has been presented by \citet{2018arXiv181107934O} using LOFAR observations.

~The application of RM Synthesis to galaxy cluster outskirts and filaments is limited by the low statistic of strong and diffuse background sources. In particular, diffuse sources would be of great interest since they probe several LOSs through the magnetized plasma.
While the JVLA may already provide a detection threshold in polarization which is low enough to detect the "tip of the iceberg" of faint background sources as well as of the diffuse emission from shocks
around filaments and clusters, a larger statistic of faint ($\leq 10-100\,\mu$Jy) sources is expected from  future radio facilities capabilities. 

In particular, future planned large radio polarisation surveys (e.g. with ASKAP, MeerKAT and the SKA-MID)   will enable the use of hundreds of sources in the background of massive galaxy clusters for background RM studies \citep[][]{2013A&A...554A.102G,2014ApJ...790..123A,2015aska.confE..92J,2015aska.confE.105G,bo15}. Moreover, statistical methods based on Bayesian inference are also being developed to allow a robust removal of the various galactic and extragalactic contribution to the observed RM \citep[][]{2016A&A...591A..13V}. The deployment of large systematic polarization surveys will also enable the use of  Fast Radio Burst 
as powerful lighthouse to study cosmic magnetic fields via RM analysis \citep[][]{2016ApJ...824..105A,va18frb,2018arXiv180810546A}. 

To date, only a few hints of the hottest ($\sim 10^{5}-10^{7}$ K) phase of the IGM, called the Warm Hot Intergalactic Medium (WHIM, see \citet{2016xnnd.confE..27N} for a recent review), have been detected by X-ray observations of cluster outskirts
\citep[e.g.][for A2744]{2015Natur.528..105E}. Complementary to this, also microwave observations of the Sunyaev-Zeldovich (SZ) effect from single objects \citep[e.g.][]{2013A&A...550A.134P,2018A&A...609A..49B} or via stacking \citep[e.g.][]{2017arXiv170910378D} proved to be effective in detecting gas in filaments connecting closely interacting clusters. 

Present or future available X-ray and SZ data provide a straightforward estimate of the gas density. Since the RM induced by a magnetized medium on polarized light depends linearly on the product of the magnetic field along the LOS ($B_{\parallel}$) and the density of the free electrons in the medium, RM observations can provide a powerful estimate of $B_\parallel$ if robust estimates of the projected thermal electron density are available.

In this contribution we use numerical simulations  to address for the first time the possibility of detecting at least the terminal part of the magnetised cosmic web, by focusing on the  observations of intracluster filaments connected to massive  galaxy clusters via the Faraday Rotation effect.  Despite the unavoidable uncertainties connected to the correlation of density and magnetic field fluctuations in the WHIM of intracluster filaments, which affect the interpretation of RM data, this approach has the advantage of being independent of the distribution of relativistic electrons on such large scale, which introduces instead other uncertainties (mostly connected to the unknown particle acceleration efficiency in such rarefied environments) in the quest for large-scale synchrotron emission from the cosmic web \citep[e.g.][]{2011JApA...32..577B,va15radio,vern17,brown17}

\subsection{Predictions of magnetic fields in filaments}

The plasma conditions of the WHIM of typical cosmic filaments are predicted to be supersonic ($\mathcal{M} \sim 1-10$), with gas accretion mostly leading to predominantly compressive turbulence \citep[][]{ry08}. In particular, under the assumption of a small-scale dynamo amplification of weak seed magnetic fields and assuming a turbulent forcing time of $\sim 10~ t_{\rm eddy}$ (where $t_{\rm eddy} \sim l_{\rm eddy}/\sigma_v$ is the eddy turnover time for an eddy with linear size $l_{\rm eddy}$ and a velocity dispersion $\sigma_v$), \citet{ry08} predicted a magnetic field strength of 

\begin{equation}
B_{\rm rms} \simeq (8 \pi~ \epsilon_{\rm turb} ~\phi)^{1/2}
\end{equation}

where $\epsilon_{\rm turb}$ is the turbulent kinetic energy density and $\phi$ is a factor that accounts for the growth of magnetic energy under the typical local conditions. \citet{ry08} estimated  $\phi \sim 0.3-0.4$ and based on the gas conditions in their simulated filaments they concluded that that $B_{\rm rms} \sim 10-10^2 ~\rm nG$ for the most massive and hot ($\geq 10^7 ~\rm K$) filaments. 

Based on the above picture, \citet{2009ApJ...705L..90C} suggested the following formula for the RM dispersion across different lines of sight crossing a filament's volume:

\begin{equation}
\sigma_{\rm RM} \sim 5 ~\rm rad/m^2 \cdot \frac{n_e}{10^{-4} cm^{-3}} \cdot ({\frac{L_{\rm fila}}{5 ~Mpc}})^{1/2} \cdot ({\frac {l_{\rm eddy}}{0.3 ~Mpc}})^{1/2} \cdot \frac{B_{\rm rms}}{100 ~\rm nG}
\label{eq:sigma_RM}
\end{equation}

where $n_e$ is the thermal electron number density, $L_{\rm fila}$ is the filament thickness and $l_{\rm eddy}$ is the integral scale of the magnetic field power spectrum. It shall be remarked that in the above picture the typical magnetic field strength, $B_{\rm rms}$, and the typical RM dispersion are reached {\it regardless} of the seed magnetic field, because in this scenario the final magnetisation is dominated by the dissipation of kinetic  into magnetic energy, in a process in which any dependence on the initial seed field is soon lost.\\

While in the above case the scales and the level of the amplified field $B_{\rm rms}$ are linked together,  direct numerical simulations of magnetic field growth in filaments have disputed the above picture, which was derived assuming an entirely solenoidal forcing of turbulence in the IGM, forced for  $\sim 10$ dynamical times \citep[][]{ry08}.  Fully MHD cosmological simulations have indeed recently  investigated the presence of small-scale magnetic dynamo amplification in cosmic filaments, reporting little-to-no evidence for volume-filling dynamo amplification \citep[e.g.][]{br05,va14mhd,2015MNRAS.453.3999M,va17cqg},
opposite to the case of  galaxy clusters simulated with the same techniques, in which case a small-scale dynamo has been observed \citep[e.g.][]{do99,va18mhd}.
The physical reason  is that  there is only a fairly limited number of dynamical times  for amplifying the field as the magnetic eddies are advected onto the neighboring clusters, with velocities of several $\sim 10^2 \rm ~km/s$, and moreover the input turbulent energy is predominantly supersonic \citep[e.g.][]{va14mhd,2016MNRAS.462..448G}. This suggests that the medium in filaments is in general an environment disfavoring the onset of efficient dynamo amplification. 

In the lack of small-scale dynamo amplification, the magnetic field amplitude is anchored to the amplitude of the seed field via compression:

\begin{equation}
 B_{\rm fila,low} \simeq B_{0} \cdot \left({\frac{n_e} {\langle n\rangle}}\right)^{\alpha_B}
\label{eq:Bcomp}
\end{equation}
where $B_0$ is the seed field, $\langle n \rangle$ is the cosmic mean (gas) density and $\alpha_B \approx 2/3$ for isotropic gas compression. Cosmic filaments are only mildly non-linear objects of the cosmic web, and their average density  is  $n_e \sim 5-10 ~ \langle n \rangle $ \citep[e.g.][]{2005MNRAS.359..272C,2006MNRAS.370..656D}, hence their average magnetic field  is only $\sim 3-5$ times larger than the original $B_0$ seed field. However, recent MHD simulations of cosmic filaments showed that the axial gas density profile of filaments is stratified at least over one decade \citep[][]{gh15}, as well as that the gas density fluctuations within filaments can extend up to 
$\sim 2-3$ decades in range, due to presence of substructures \citep[][]{2016MNRAS.462..448G}. Hence while Eq.~\ref{eq:Bcomp} gives a  lower limit on the average magnetic fields in filaments for a given primordial magnetic seed, the internal distribution of magnetic field fluctuations can be as large  as $\rm B_{\rm fila,high} \sim 10^2-10^3 ~B_0$ in simulations \citep[][]{2016MNRAS.462..448G}.  If this also happens in the real Universe,
 we should expect that the distribution of RMs from filaments can stretch over $\sim 4-5$ orders of magnitude, since density and magnetic field fluctuations are well correlated in the compressive regime, which  would likely bias detections of RM  towards the highest value of the distribution.\\

A scenario complementary to the "primordial" one  is the one in which the seeding of magnetic fields in large-scale structures is entirely due to the 
"pollution" of magnetic fields by 
 active galactic nuclei and galactic activities \citep[e.g.][]{donn09,xu09}. In this case, less correlation should be expected between density and magnetic field fluctuations, yet the range of magnetic field values in filaments (and hence of RMs) is expected to be much more extended  
than in a primordial scenario, owing to the dilution of ejected magnetic fields away from sources, as well as by the expected drop in the number density of sources moving into the less dense Universe  \citep[e.g.][]{2015MNRAS.453.3999M,va17cqg}.\\

\bigskip

\section{Methods \& Materials}

\subsection{Simulations of extragalactic magnetic fields}

We simulated the formation of massive galaxy clusters
using a customised version of the cosmological grid code {\enzo} \cite[][]{enzo14}.  We  used the Dedner formulation of MHD equations \citep[][]{wa09} and used adaptive mesh refinement (AMR) to  increase the dynamical resolution \citep[e.g.][]{xu09}.

In this work, we mostly focus on {\it non-radiative} cosmological simulations that include only the effects of cosmic expansion, gravity and (magneto)hydrodynamics. However, in the discussion (Sec~.\ref{subsec:AGN}) we also include for completeness simulations with cooling and feedback by active galactic nuclei. 

Each cluster forms in a volume of (260 Mpc)$^3$ (comoving), and is simulated starting from a root grid $256^3$ cells and using $256^3$ dark matter particles. The initial density perturbation field is taken from a suite of existing cluster simulations \citep[e.g.][and other works derived from this]{va10kp}. The innermost  $\sim$ 25 Mpc$^3$ volume, centred on where each cluster forms, has been further refined 5 times ($2^5$)using AMR. Mesh refinements are initiated wherever the gas density is $\geq 1\%$ higher than its surroundings. This give us a maximum spatial resolution of $\Delta x_{\rm max} \approx 31 ~\rm kpc$. The mass resolution for dark matter particle in the high resolution region is $m_{\rm DM}=9.1 \cdot 10^{10} M_{\odot}$ for all clusters. 

The assumed cosmology in this work is a $\Lambda$CDM model with: $h = 0.72$, $\Omega_{\mathrm{M}} = 0.258$, $\Omega_{\mathrm{b}}=0.0441$ and  $\Omega_{\Lambda} = 0.742$. 

In this work we focus on three massive ($M_{\rm 100}\sim 10^{15} M_{\odot}$) simulated galaxy clusters, drawn from a larger sample: 
 a) cluster "e1", with a virial mass of $M_{\rm 100}=1.12 \cdot 10^{15} M_{\odot}$ and a virial radius of $2.67$ Mpc, which was interested by a major merger at $z \approx 0.1$; b) cluster "e14",  with a virial mass of $M_{\rm 100}=1.00 \cdot 10^{15} M_{\odot}$ and a virial radius of $2.60$ Mpc, which is in a fairly relaxed dynamical state by z=0;  c) cluster "e18b", with a virial mass of $M_{\rm 100}=1.37 \cdot 10^{15} M_{\odot}$ and a virial radius of $2.80$ Mpc, which was interested by a major merger at $z \approx 0.5$ and still is in a perturbed dynamical state by $z=0$. 

\begin{figure}[H]
\centering
\includegraphics[width=0.6\textwidth]{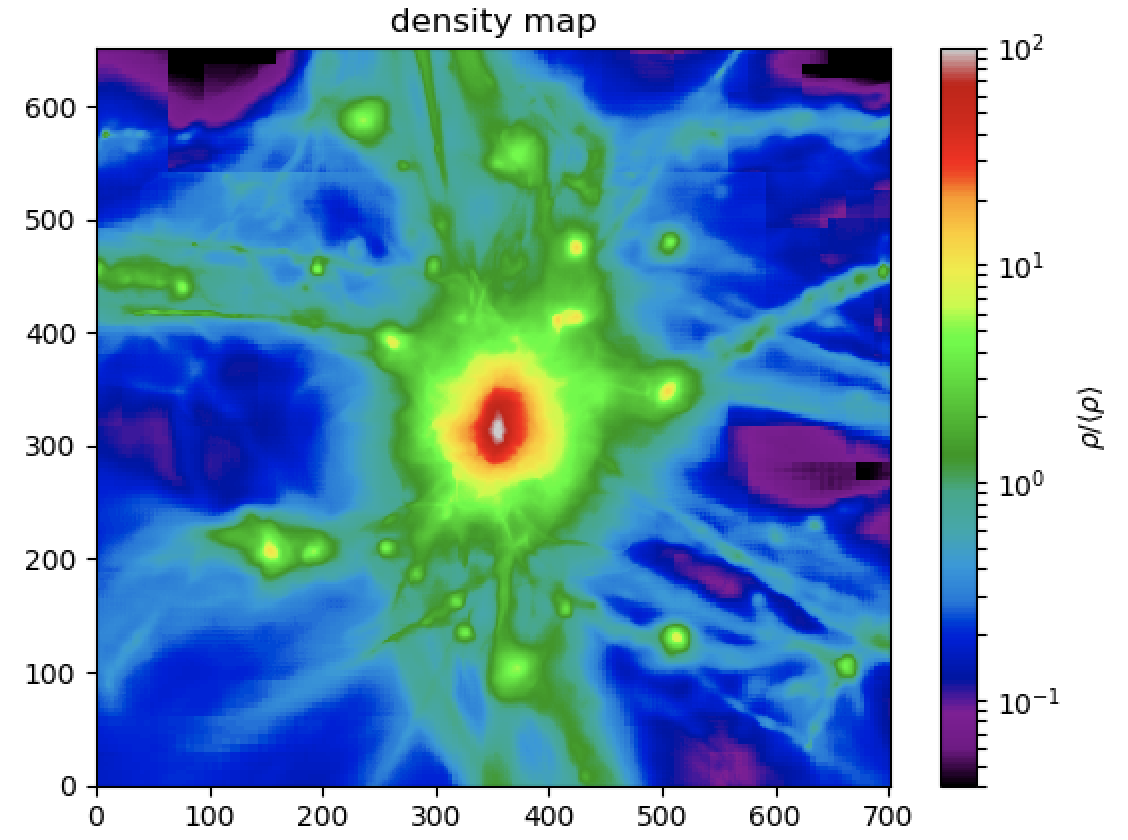}
\includegraphics[width=0.7\textwidth]{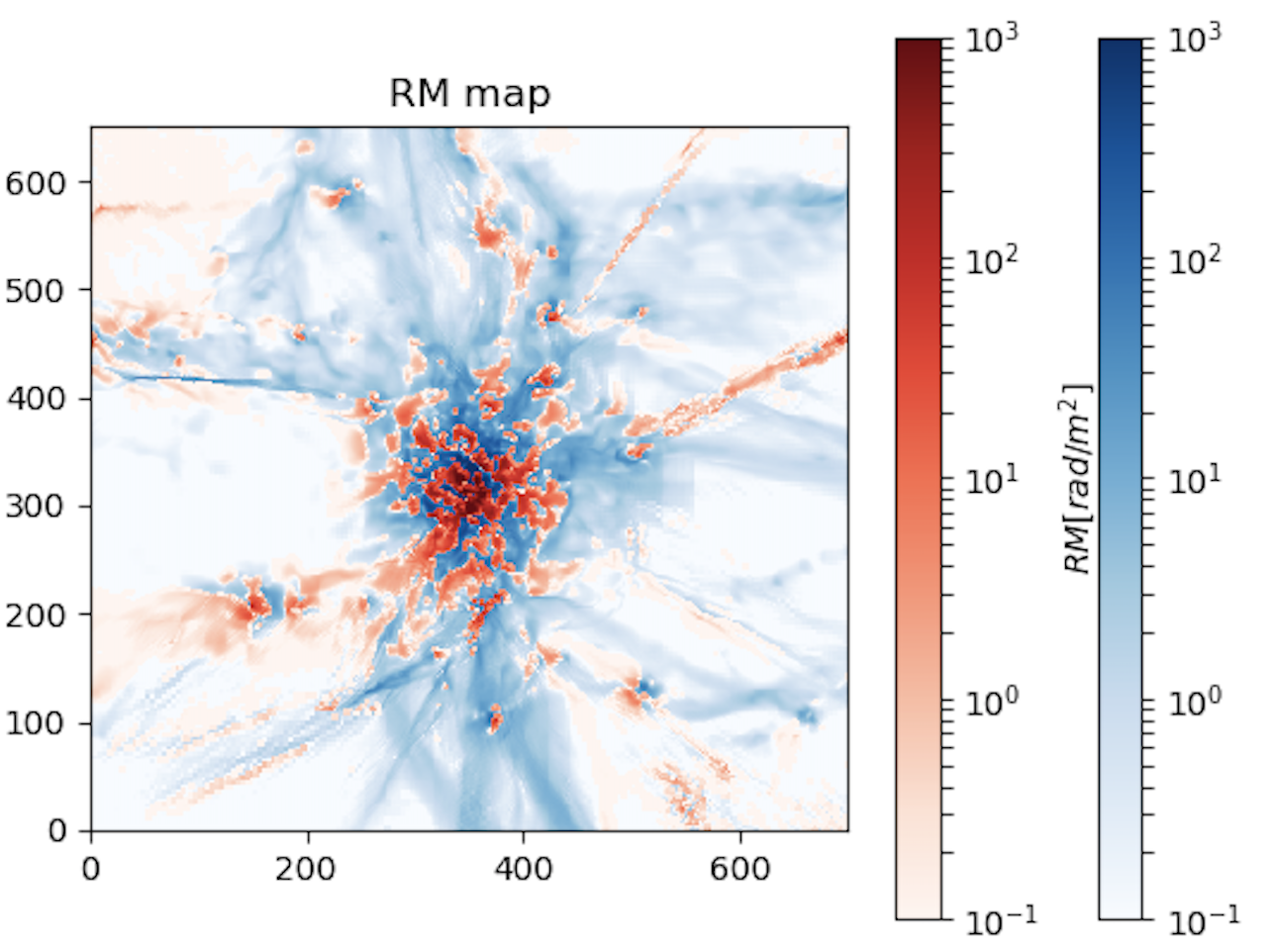}
\caption{Projected gas density (left) and RM (right) for our cluster e18b at $z=0$,  for the non-radiative run with primordial magnetic fields. Each box has sides $22 \times 20 ~\rm Mpc^2 $. In the RM map, we show with red colors pixels with $\rm RM > 0$ and with blue colors the pixels with $\rm RM <  0$.}
\label{fig:fig1}
\end{figure}

Our baseline model for the magnetic field in clusters is a simplistic "primordial" seeding scenario, in which we initialised the magnetic field to a uniform value $B_0$ across the entire computational domain, along each coordinate axis \citep[][]{wi17,va18mhd}. The initial magnetic seed field of $0.1 ~\rm nG$ (comoving) is chosen to be below the upper limits from the analysis of the CMB \citep[e.g.][]{sub15}, as well because with this initial magnetic field strength our simulations are able to produce a reasonable match to observed radio relic power \citep[][]{wi17} as well as to the observed Faraday Rotation profile for the Coma cluster \citep[][]{va18mhd}. 
An example of the projected gas density and RM for one of our simulated cluster is given in Fig.\ref{fig:fig1}, which shows the volume around each clusters that will be subject of our filament analysis. We notice already that the visible part of filaments connected to our objects is only a small fraction of the much more extended (and more rarefied) length of typical cosmic filaments, which is of several tens of $\rm Mpc$ in the cosmic volume \citep[e.g.][]{gh15}.

These simulations are non-radiative and there are no sources of thermal, kinetic or magnetic feedback. In order to bracket uncertainties, we will also test in Sec.~\ref{subsec:AGN} resimulations of the same clusters using an alternative scenario in which magnetic fields in the same objects have been seeded by past activity of AGNs.

\subsection{Filament selection and properties} \label{subsec:filament_selection}

Starting from the small sample presented above, we pre-selected filaments in each cluster and analyzed their projected properties along the three coordinate axis, in order to have preliminary set of 9 independent targets for our mock observation of RM in filaments. 

For every 2-dimensional radial shell (with a fixed radial bin of $\Delta x = 31 \rm kpc$) and starting from $R_{\rm 100}$ we selected pixels with a projected density in the $66-98 \%$ of the density distribution, which efficiently tracks the gas which is structured in filaments but is not clumped (by self-gravity and over-cooling) into dense substructures.  Additionally, we limited our analysis to the $T \geq 10^6 ~\rm K$ projected temperature, in order to include only gas that can be potentially associated to soft X-ray emission. This procedure is designed to broadly mimic the selection based on X-ray or optical observations, which will become available in the next decade due to the combination of large surveys (e.g. Euclid, eRosita and more in the future, Athena). 

The filaments were named with the first three characters referring to the simulated cluster from which they were taken (e01, e14 or e18); a letter telling the axis along which the cluster was observed in order to retrieve different realizations (X, Y or Z); one or two letters telling the filament direction on the sky with respect to the cluster (N=North, S=South, E=East, W=West). Although automatic algorithms for the detection of cosmic filaments in 3-dimension have been developed also for our simulations \citep[][]{gh15}, here in the further analysis of data for simplicity we proceeded to the {\it visual} identification of single filaments around each cluster. The final dataset used for the following analysis consists in 29 filaments in total.

\subsection{Mock Rotation Measure observations}

Our predictions assume a  uniform random distribution of polarised 
radio sources in the background of our clusters.
For all 29 filaments in our sample, we extracted 
 an increasing number ($\rm N_S=\{ 5,10,15,20,25,40,65,100 \} $) of sources at random locations, and computed the statistical distribution of RM across the Faraday screen produced by each cluster and its environment (within a $\approx 25^3 ~\rm Mpc^3$ volume). For each run, we considered an equal number of sources in a "control field" (i.e. a field where there are no filaments or galaxy groups) as well as in a "filament" field (selected as above). \\

Beside the effect of external RM from the cosmic web, for each mock observation we additionally included:
\begin{itemize}
\item A fixed contribution to RM from the Galactic foreground. We restricted ourselves to targets at high galactic latitude ($\geq 80^{\circ}$), for which the RM contribution is in general of $|RM_{\rm Gal}| \leq 10 ~\rm rad/m^2$ \citep[][]{2015A&A...575A.118O}. In particular, we assumed here for simplicity a fixed $+6.0 ~\rm rad/m^2$ contribution to each field, noticing that this contribution should in general be the easiest to tell apart in real observations, because the extent of the typical size of filaments around galaxy clusters we consider here is 
$\leq 0.1-0.5^{\circ}$, i.e. much smaller than the typical angular scales of variations of the Galactic foreground. For example, based on Eq.~20 in  \citet{2015ApJ...815...49A}, we can estimate a typical RM fluctuation of $\leq 0.5 \rm ~rad/m^2$ across $0.5^{\circ}$ from the Galactic foreground. 
\item { a residual contribution to the RM, RM$_{\rm res}$, which includes an internal contribution to each background source RM$_{\rm src}$; a contribution from other extragalactic sources as intervening MgII absorbers RM$_{\rm MgII}$ \citep[][]{2013MNRAS.434.3566J}; a residual RM after Galactic foreground subtraction which can be present on scales smaller than the one used to fit the Galactic contribute RM$_{\rm MW,res}$. This value has been estimated by \citet{2010MNRAS.409L..99S} to be normally distributed as $\sigma_{\rm res} \leq 6$~rad/m$^2$ and has been confirmed by \citet[$\sigma_{\rm ERS}$ in their work]{banfield14}. We follow their treatment and keep RM$_{\rm src}$, RM$_{\rm MgII}$ and RM$_{\rm MW,res}$ together in the one factor RM$_{\rm res}$.} However \citet{banfield14} note that the value is dependent on the background source population. At 1.4~GHz, the WISE-AGN population defined in \citet{2011AAS...21832810J} biases the estimate of $\sigma_{\rm res}$ and shows a larger $\sigma_{\rm res}=12 \pm 0.2$~rad/m$^2$. We thus consider the latter as a more conservative case, while we considered $\sigma_{\rm res}=6$~rad/m$^2$ as a standard case based on literature works \citep{2010MNRAS.409L..99S,banfield14} that can be optimized e.g. using only star-forming nearby galaxies for background studies. 

We thus randomly draw the RM$_{\rm res}$ value from a Gaussian distribution with standard deviation $\sigma_{\rm res}=12$ or $6~\rm rad/m^2$, depending on the assumed background source population. 
\item The estimated error to RM $\delta_{RM}=\sqrt{3}/({\rm SNR}_P \, \Delta\lambda_{\rm max}^2)$ \citep{2005A&A...441.1217B,2014ApJ...785...45R} where SNR$_P$ is the signal-to-noise ratio of the source in the polarized image and $\Delta\lambda_{\rm max}^2$ is the difference between the largest and smallest observed $\lambda^2$.  Also for the latter we considered two possible different values: a) JVLA-like observations in which $\delta_{\rm RM}=8 ~\rm rad/m^2$ (assuming wide total bandwidth $\Delta\nu\simeq1$GHz, L-band observations of background sources with SNR$_P>3$); b) SKA-MID-like observations in which $\delta_{\rm RM}=1 ~\rm rad/m^2$ which corresponds to current estimates for SKA-MID performances \citep[e.g.][]{2015aska.confE.105G}. 

\end{itemize}

The simulated statistics are important to assess the crucial improvements in the significance of detections, as a function of the number of detected RMs (N$_S$) in the field and in filaments.

\subsection{Non-parametric tests of RM distributions}

We tested the null hypothesis that two random sets of RMs in filament and control field to belong to the same parental distribution.  For each random run, the two sets of RMs are assembled by extracting N$_S$ random values from the simulated distribution of RMs inside and outside the filamentary projected environment respectively, and added with noise, as defined in the previous section~\ref{subsec:filament_selection}.

~ In detail, we performed the random extraction of 1000 trials for an increasing number of $\rm N_S$ sources, and computed the distribution of p-values from the Mann-Whitney (M-W) U test \citep[][]{mw47}. We reject the null hypothesis whenever the p-value of the test is lower than or equal to $\alpha\equiv0.05$, where $\alpha$ is the significance level.  The Mann-Whitney U test has the advantage to test the equivalent null hypothesis that it is equally likely that a randomly selected value from one sample will be less than or greater than a randomly selected value from a second sample, which is precisely what we want to determine. This test does not require any assumption on the two compared distributions (non-parametric test).\\
Compared to the other more widely used non-parametric test: the Kolmogorov-Smirnov (K-S) D test, the M-W U test is known to be more reliable for small samples and therefore best suited for our analysis. \\
High values of the rejection fraction given in the following figures will indicate high chance to detect a RM excess scatter in the filament set, with respect to the control field. Our fiducial rejection fraction threshold for detection is set to 0.2. Though it gives a still low chance probability (1/5), rejection fractions above this level are shown to actually improve with an increasing number of polarized background sources $N_S$. Lower rejection fractions are instead mostly random values.

%%%%%%%%%%%%%%%%%%%%%%%%%%%%%%%%%%%%%%%%%%
\section{Results}

\subsection{Detectability of intracluster filaments using RM}

We can now assess the chances of  detecting the magnetic cosmic web attached to massive galaxy clusters under realistic observing conditions, by focusing on the challenge of significantly distinguish the excess RMs in filaments compared to control fields. 
In this mock observing procedure, we assume to know the approximate location of filaments 
in the cluster neighborhood based on the X-ray or optical data \citep[e.g.][]{2015Natur.528..105E,2018arXiv180908241C}, and studied how many detected RMs will be necessary 
to statistically distinguish filaments from control fields, based on the outcome of the M-W test (averaged over 1000 independent realizations for each number of sources). \\

\begin{figure}[H]
\centering
\includegraphics[width=0.49\textwidth]{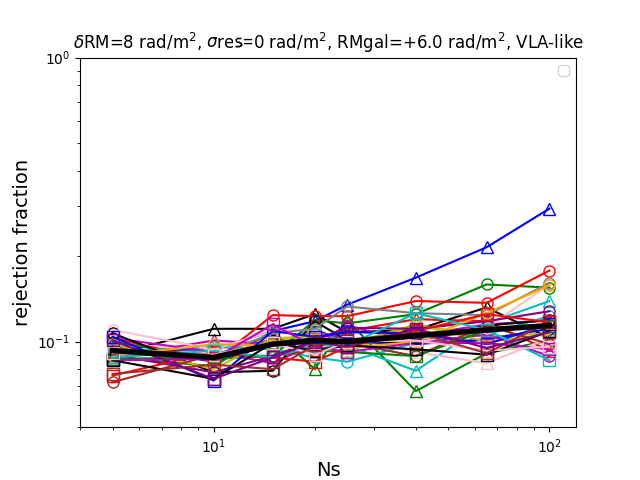}
\includegraphics[width=0.49\textwidth]{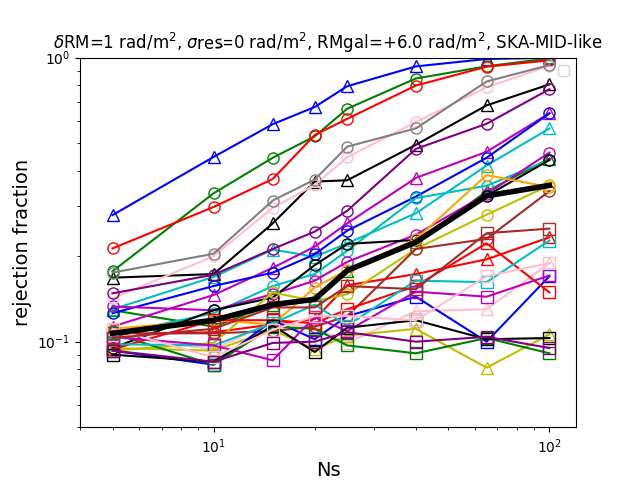}
\caption{Rejection fraction for an increasing number of detected RMs calculated for all the simulated filaments in the primordial seeding scenario,  for an idealized scenario in which we set $\sigma_{\rm res}$ to 0~rad/m$^2$ for the residual RM on background sources. Different colors mark different filaments, different marks point to different simulated clusters, while the solid black line show the median for fixed N$_S$. Left: JVLA-like parameters; Right: SKA-MID-like parameters. Color and mark codes of filaments are conserved through all the plots in this work.}
\label{fig:rej_frac_RMers0}
\end{figure}

First, we show in Fig.~\ref{fig:rej_frac_RMers0} the  rejection fraction of all the filaments in the sample for both a JVLA-like observation (left panels) and a SKA-MID-like one (right panels), as a  function of  N$_S$ and in the ideal case in which there is no contamination from the residual RM on background sources ($\sigma_{\rm res}=0~\rm rad/m^2$). In this case, the big limiter for the efficient detection of filaments compared to control fields is the sensitivity of the radio telescope. Even without the contamination from background and intervening sources, a JVLA-like observation will be able to significantly detect only a handful of objects only by observing N$_S \geq 50$ sources. On the other hand, with the $\sim 8$ times increased sensitivity of the SKA-MID we observe a remarkable improvements in the simulated detection rate, with the majority of objects in the sample being fully (or marginally) detectable for 
a large number of sources. Even with $\sim 10$ sources per filament, a half of the objects in our sample should be statistically detectable against control fields. 

However, the situation dramatically changes as soon as the additional contribution from the residual RM is included in the analysis.

\begin{figure}[H]
\centering
\includegraphics[width=0.49\textwidth]{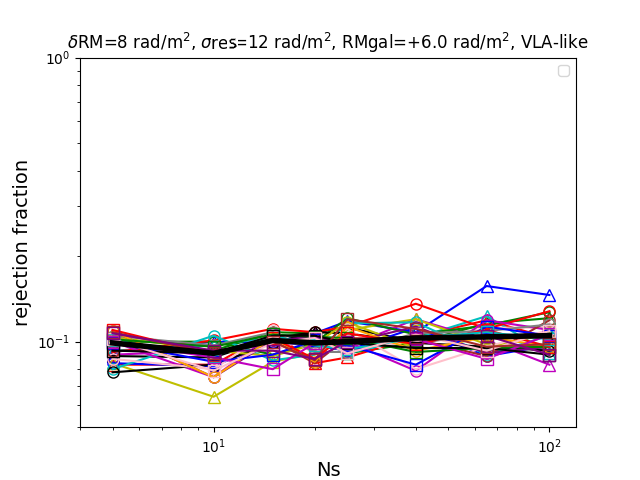}
\includegraphics[width=0.49\textwidth]{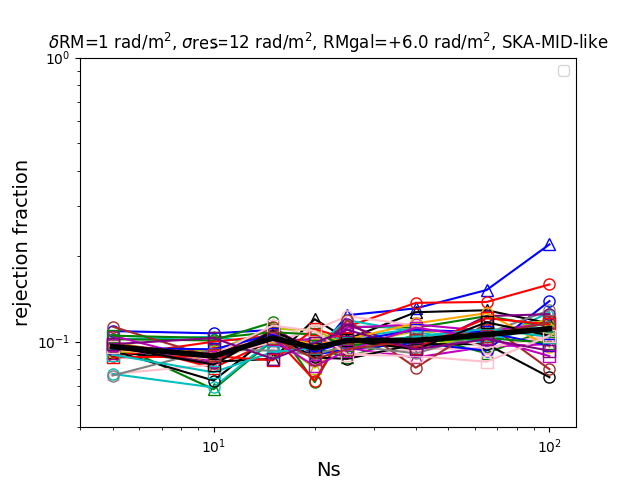}
\includegraphics[width=0.49\textwidth]{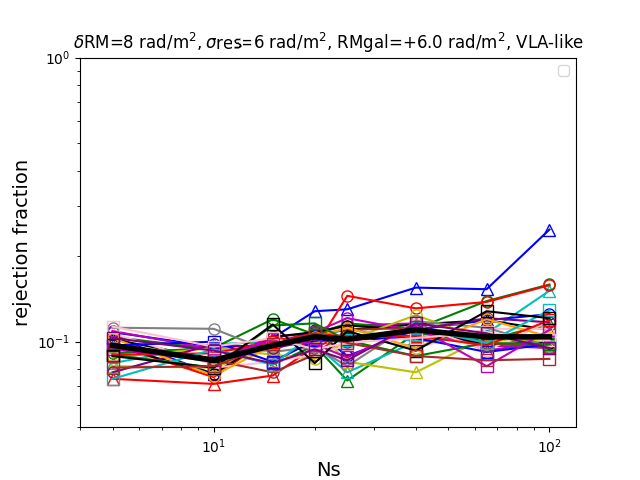}
\includegraphics[width=0.49\textwidth]{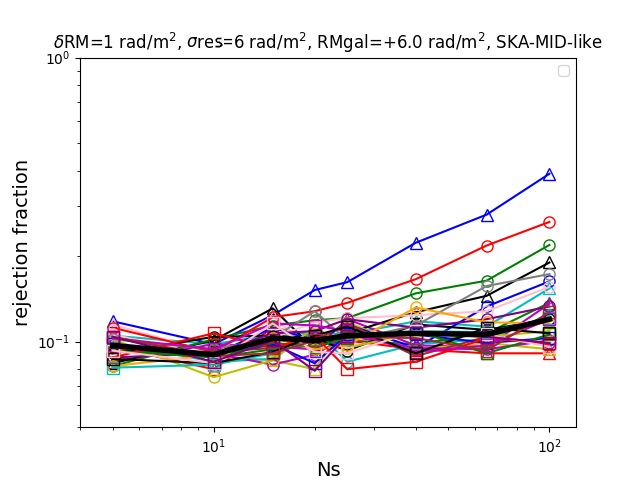}
\caption{Rejection fraction for an increasing number of detected RM as in Fig.~\ref{fig:rej_frac_RMers0}, but by assuming $\sigma_{\rm res}=12$~rad/m$^2$ for the residual RM of sources (top panels), while the lower panels we assumed $\sigma_{\rm res}=6$~rad/m$^2$.}
\label{fig:rej_frac}
\end{figure}

In Fig.~\ref{fig:rej_frac} we plot the rejection fraction of all the filaments in the sample for both a JVLA-like observation (left panels) and a SKA-MID-like one (right panels), as function of  N$_S$. 

For the magnetic field model considered here, it seems challenging for any JVLA-like observation to robustly distinguish the RM distributions of filament and control field sets, also with the unrealistically large number of N$_S=100$ sources per single object.  This is due to the low RM contribute of the Faraday screen (i.e. the magnetized plasma in the filament) when compared with the scatter of the assumed distribution of RM produced both internally and along the line of sight outside the filament ($\sigma_{\rm res}$). 
\citet{banfield14} attribute the RM scatter along the line of sight as a contribute of both internal RM and intervening magnetized substructures and show that it can vary with the background source population. By considering the 
possibility of selecting only background sources with $6$~rad/m$^2$ (Fig.~\ref{fig:rej_frac}, lower panels) our test show that a tentative detection of a few prominent filaments becomes possible for N$_S \geq 50$ sources. 

We also notice that even in the most optimistic situation considered here, i.e. a survey with SKA-MID and a low contribution from residual RM on sources, the statistical detection of intracluster filaments will be feasible only for a small fraction of objects.

All our results can be summarized as follows: 
\begin{itemize}
\item On average, with our procedure we estimate that galaxy clusters have $3.3 \pm 0.9$ projected filaments which can be identified by X-ray inspection. This result is consistent with the results by \citet{2005MNRAS.359..272C},  who found that $\sim 80\%$ of clusters have 1 to 4 projected connections between them.
\item Filaments selected in our procedure are on average $\sim 10-50$ times denser that the smooth environment around galaxy clusters, have a mean temperature of $T \sim 1-5 \cdot 10^{7} ~\rm K$ and an average magnetic field of $B_{\rm rms} \sim 10-50 ~\rm nG$ (see Sec.~\ref{subsec:detect}). 
\item  The typical RM in filaments is in the range $\sim 0.2-2 ~\rm rad/m^2$ for the  primordial seeding scenario considered here, i.e. 
a factor $\sim 10^2$ larger than the average RM distribution in our control fields for the detectable ones, $\sim 30$ times larger for filaments in general. However, the distributions of RMs can reach up to $\sim 10$~rad/m$^2$ in a few $\%$ of the cells, and the chances of confirming the presence of magnetic fields in filaments rely on the detection of such rare fluctuations. 

\item The rejection fraction has been fitted with a power-law trend with respect to the number of detected sources N$_S$. Considering  the 7 filaments with the highest rejection fraction in the most favorable case (i.e. extragalactic residual RM noise is small or absent) , Fig.~\ref{fig:rej_frac_RMers0}, right panel),  the best fit gives rf$\propto$ N$_S^{0.55 \pm 0.05}$ before saturation. To this end, just the filaments showing an improved rejection fraction above detection threshold even for small samples  ($N_S=5,10,15$) were considered. Including the other filaments would affect the trend with random low rejection rates.

\item Limited to the most favorable objects and for low contribution from residual RM on the sources,  the increased sensitivity that will be provided by SKA-MID compared to 
JVLA-like observations improves the rejection fractions distribution by a factor {$\sim 3(1.5)$} at N$_S=100 \, (20)$, while the number of observations with a rejection fraction larger than 0.5 increases from 0(0) to 9(3) over 29 objects.
\item The actual limiting factor for the detection of filaments is the extragalactic residual RM scatter $\sigma_{\rm res}$,  more than the number of detected sources N$_S$ throughout the field. Going from $\sigma_{\rm res}=0$~rad/m$^2$ to $\sigma_{\rm res}=6 (12)$~rad/m$^2$ the rejection fraction median drops down by a factor 3(3.3) even for the SKA-MID-like observation with 100 sources per set, and the observations with a rejection fraction larger than 0.5 falls from 31$\%$ to 0$\%$ even for this large N$_S$ value.
\end{itemize}

Given the results above, how do we set an observation able to retrieve a sufficient number of polarized background sources?\\
The current estimate for the surface density of polarized sources $n_s$ at highest resolution (1.6") at $1.4$~GHz is given by \citet{2014ApJ...785...45R} who find $n_s = 45 (P/30\mu{\rm Jy})^{-0.6}\,{\rm deg}^2$, determined in the GOODS-N field down to a sensitivity of $14.5\,\mu$Jy. 
Assuming a physical area $A_{\rm fil}$ and a distance $z$ for the target, we can estimate the necessary sensitivity $P$ in a similar observation (1.6" resolution at 1.4GHz) to sample a typical intracluster filament and detect a number $N_s$ of polarized sources

\begin{equation}
P \simeq 3\mu{\rm Jy} \left(\frac{A_{\rm fil}}{25{\rm Mpc}^2}\right)^{\frac{5}{3}} \left(\frac{N_s}{100}\right)^{-\frac{5}{3}} \left(\frac{z}{0.1}\right)^{-3.03}
\end{equation}

where we assumed the following scaling between physical and angular size of an object with redshift $z$ to be
$1.857 [{\rm kpc/"}] (z/0.01)^{0.91}$ (this relation is correct within a 7\% relative error for the redshift range $0.01 \leq z\ \leq 0.3$).
We remark that the estimate of $P$ given here is valid, strictly speaking, only for a $\simeq (1.5")^2$ synthesized beam. Determining the sensitivity at lower resolutions than the one given is not a trivial task. Beam depolarization effects are in fact introduced and they can be properly taken into account just by modeling the polarization structure of the background source populations.\\
The SKA-MID survey is planned to reach $0.09 ~\mu{\rm Jy}$ in 1000 hrs at $1.4 ~\rm GHz$ frequency and with a $\sim 3$ times finer resolution ($\simeq 0.5$")\cite{2015aska.confE..95B}. A typical intracluster filament would thus cover $\simeq 0.55\,{\rm deg}^2$ in the sky, which is approximately the putative SKA-MID field-of-view ($\simeq 0.49\,{\rm deg}^2$).
A JVLA obseravation with similar settings (1GHz total bandwidth centered at 1.4GHz, A-configuration array with resolution 1.5", 40\% flagged data) would cover the target with a 4 pointings mosaic, requiring 68hrs of total observing time to reach the required sensitivity of $3\mu$Jy per beam (17h per single pointing). \\

In summary, while a $\sim  \times 8$ increase in RM sensitivity with the SKA-MID will in principle allow the detection of a ten-fold larger amount of filaments surrounding galaxy clusters, in practice the unavoidable disturbance by the intrinsic RM noise of polarized sources will dramatically limit the number of objects for which detections of RM can be made statistically significant compared to control fields. 

%%%%%%%%%%%%%%%%%%%%%%%%%%%%%%%%%%%%%%%%%%

%%%%%%%%%%%%%%%%%%%%%%%%%%%%%%%%%%%%%%%%%%
\section{Discussion} \label{sec:discussion}
\subsection{Properties of most detectable filaments}
\label{subsec:detect}

What makes a filament more likely to be significantly  detectable compared to the majority? 

\begin{figure}[H]
\centering
\includegraphics[width=0.99\textwidth]{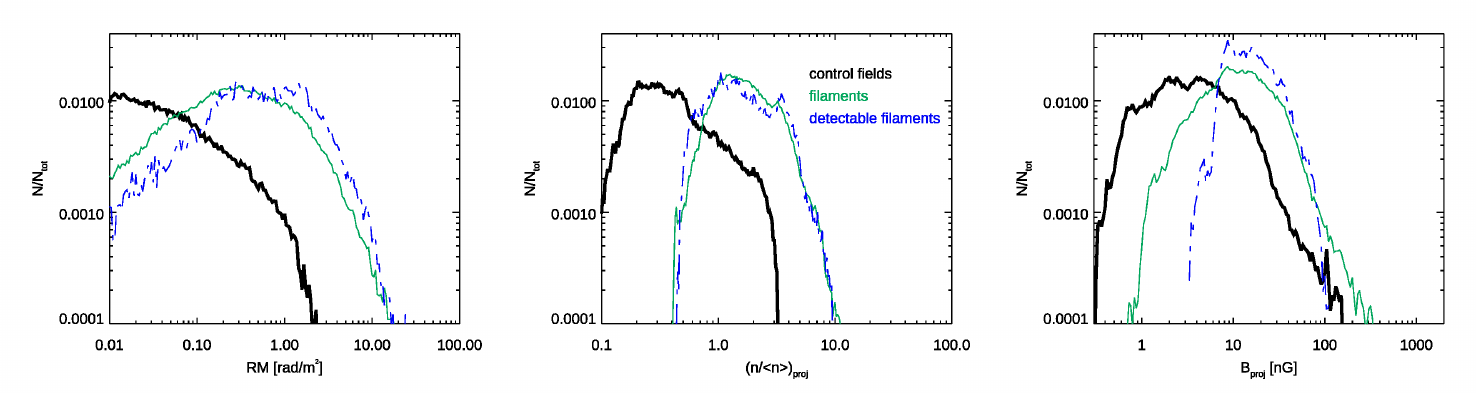}
\caption{Distributions of RMs, projected mean gas density and  mean magnetic field strength for all control fields and filaments considered in our datasets (primordial model, including all analyzed lines of sight). The additional dot-dashed lines show the distributions of the same fields limited to  the 4 most detectable filaments in Fig.~\ref{fig:rej_frac}. }
\label{fig:pdf_primordial}
\end{figure}

To address this question we computed the distribution of the projected gas density and mean magnetic field for the entire distribution of pixels in filaments or control fields, and contrasted this with the distribution from a subset of the 4 most detectable filaments (Fig.\ref{fig:pdf_primordial}), based on the previous Section. 

Somewhat surprisingly, both the  distribution of projected gas density and of projected magnetic field strength do not significantly differ when we compare lines of sight crossing detectable filaments with the rest of the population (while, of course, lines of sight crossing filaments are significantly denser and more magnetised than  lines of sight crossing pixels in the control fields).  However, the RM also depends on the distribution of scales in the magnetic field (as in Eq.~\ref{eq:sigma_RM}).

\begin{figure}[H]
\centering
\includegraphics[width=0.79\textwidth]{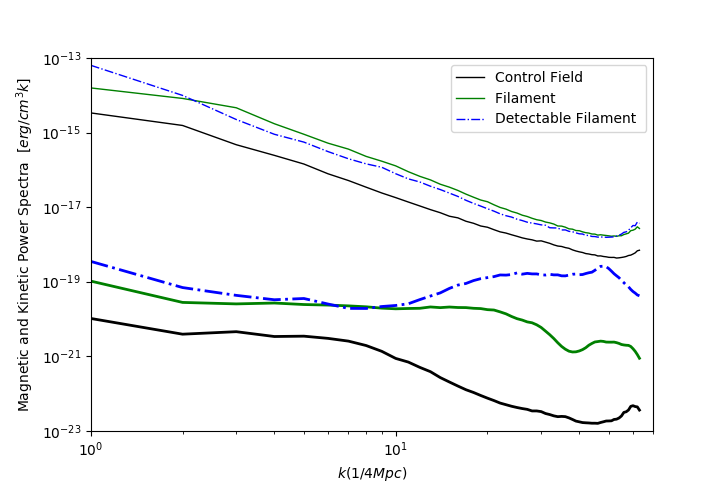}
\caption{The upper thin lines show the 3-dimensional power spectra for the density weighted velocity field ($\rho^{1/2} v$) and the lower thick lines show the 3-dimensional magnetic power spectra in three sub-volumes of the cluster e18b.}
\label{fig:spectra}
\end{figure}

We thus computed  the power spectra of magnetic fields in different 3-dimensional sub-volumes in the field of our cluster e18b, motivated by recent simulations by our group, in which the signature of magnetic dynamo in the innermost cluster regions clearly stems in power spectra \citep{2018arXiv181008009D}.
in particular we selected a cubic  $4^3 \rm ~Mpc^3$  box coincident with a clearly detectable filament connected to the cluster (right of the cluster centre in Fig.1), a similar box coincident with an undetectable filament (left of the cluster centre), and a cubic volume on an empty "control field" located at the cluster virial radius of e18b.  The 3-dimensional power spectra of magnetic field and of the density-weighted velocity field ( $\rho^{1/2} ~\vec{v}$  {\footnote{We notice that the density weighting in the velocity spectra ensures that the magnetic and velocity spectra have the same units and can be quantitatively compared, as in \citet{va18mhd}.}} ) were computed with a standard  Fast Fourier Transform (FFT) algorithm, similar to our previous work \citep[e.g.][]{va14mhd}, and are shown in Fig.~\ref{fig:spectra}.  

Regions containing filaments show a $\sim 5-10$ larger kinetic energy budget at all scales, while the magnetic field energy is even $10^3-10^4$ larger than the control volume for 
 $k \geq 10$ ($\leq 400 \rm ~kpc$) scales. On such scales, the magnetic energy in the detectable filament is a few percent of the kinetic energy, while the it is $\leq 10^{-5}$ of the kinetic energy in the undetectable filament. The visual inspection further suggests that  the 
sub-volume around the detectable filament contains more gas substructures, which are likely responsible for enhanced density fluctuations and for the mixing of magnetic field lines on small-scales. While the clumpiest part of this volume is excised from our RM analysis (see Sec.\ref{subsec:filament_selection}), a higher degree of structures within the filament implies that the environment has surely been subject to a higher dynamical activity in the recent past, which significantly boosted the magnetic field beyond compression. 

In summary, the most promising filament targets seem to be characterized by a higher level of substructures, associated with ongoing clump accretion, which enhances the tangling of magnetic field lines on $\leq 400 ~\rm kpc$ scales, boosting the overall RM signal.

\subsection{Alternative models of magnetic fields in filaments}
\label{subsec:AGN}

\begin{figure}[H]
\centering
\includegraphics[width=0.55\textwidth]{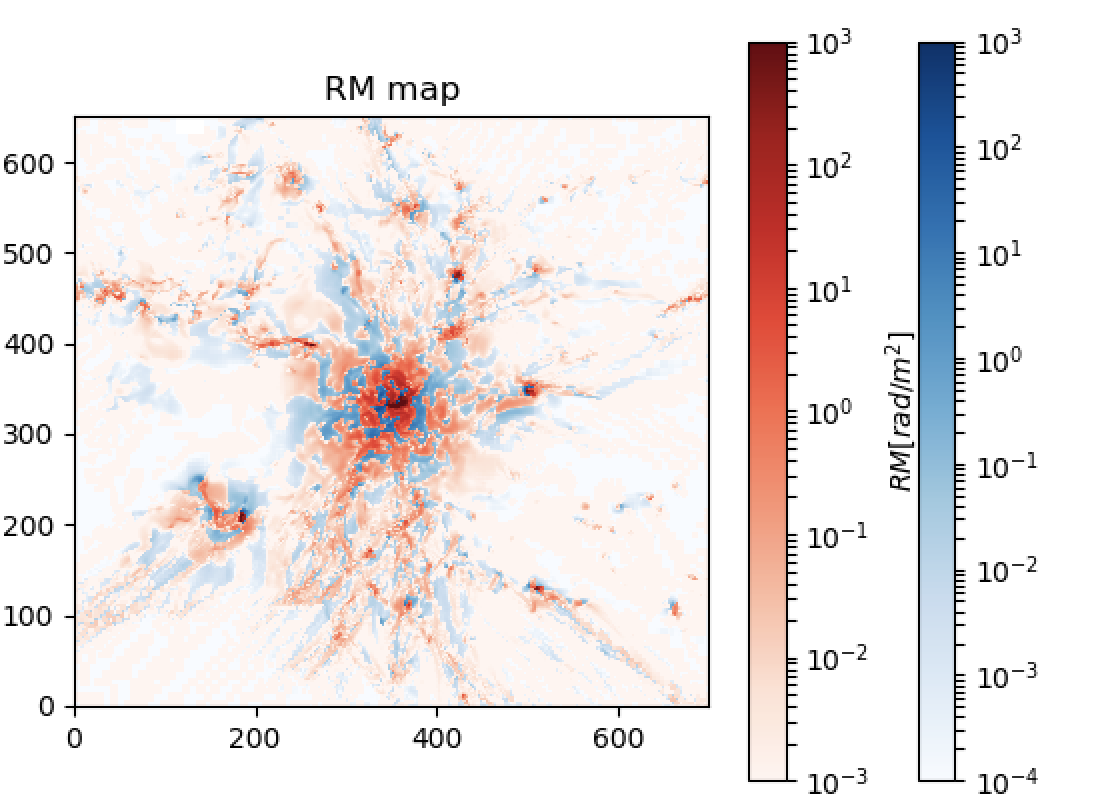}
\includegraphics[width=0.44\textwidth]{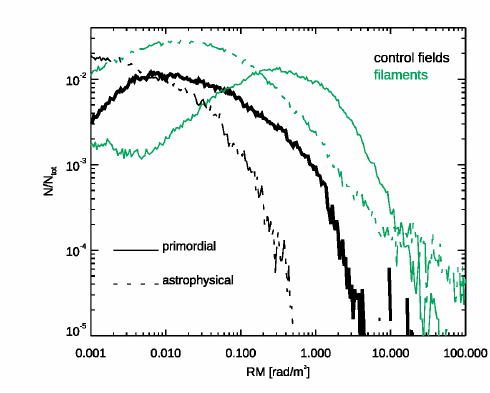}
\caption{Left panel: RM distribution  for our e18b cluster at $z=0$, simulated with the  cooling and feedback model, in which magnetic fields are injected by AGN activity. The box as the meaning of colors is as in Fig.1.  Right panel: distribution of RMs for all filament and control fields, for the primordial and the astrophysical model.}
\label{fig:e18b_agn}
\end{figure}

In order to bracket uncertainties related to the (unknown) origin of extragalactic magnetic fields and on the details of gas physics, we first contrasted our baseline "primordial" scenario with a second "astrophysical" scenario, in which we employed radiative simulations with a simple prescription for feedback from active galactic nuclei (AGN). In such runs the gas loses energy at run-time assuming equilibrium cooling for a primordial chemical composition, and the launching of bipolar thermal jets from simulated AGNs, which also deliver a fixed fraction ($\approx 1 \%$) of the feedback energy into magnetic energy (see \citet{va17cqg} for more details). We show in Fig.~\ref{fig:e18b_agn} the example of the RM for cluster e18b at $z=0$, which well illustrates how in this second simulation the number of gas substructures is increased, while the level of RM in the diffuse WHIM is smaller compared to the primordial case, owing to the dilution of local sources of magnetisation outside of clusters. The right panel in the same Figure also shows the overall change in the distribution of RM in control fields and in filaments for the entire dataset in the two cases. 

\begin{figure}[H]
\centering
\includegraphics[width=0.49\textwidth]{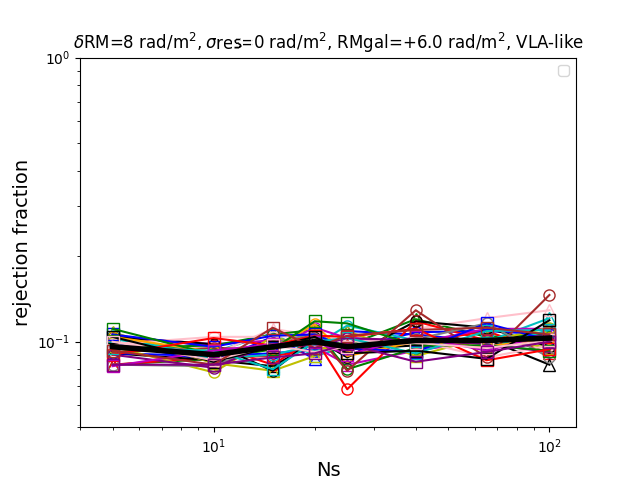}
\includegraphics[width=0.49\textwidth]{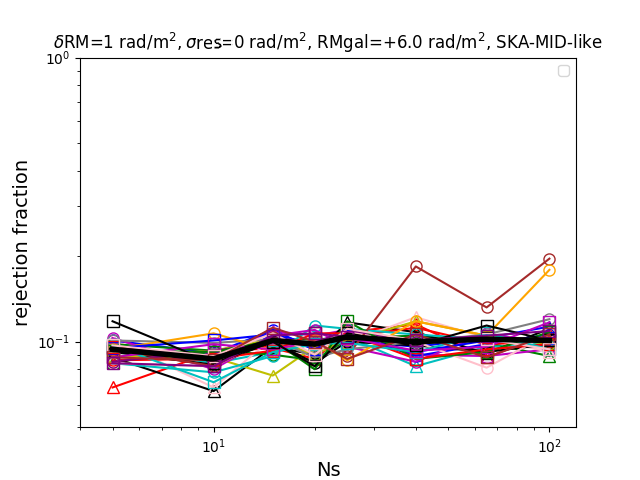}
\caption{Same as Fig.~\ref{fig:rej_frac_RMers0} for the AGN seeding scenario. }
\label{fig:rej_frac_AGN}
\end{figure}

Although in the AGN seeding model we measure a tail of  very high RM values (RM$>20$~rad/m$^2$) in the ICM and filaments, and lower RM values in the control field (which produces a larger contrast between them), telling the two models apart based on the statistical analysis of observable RMs becomes more difficult (see Fig.~\ref{fig:rej_frac_AGN}).  Virtually no filaments would be detectable in this case, not even by considering no contamination from residual RM ($\sigma_{\rm res}=0$~rad/m$^2$), and even for a SKA-MID-like observation (Fig.~\ref{fig:rej_frac_AGN}, right panel). 

However, a possible way to investigate the AGN seeding scenario, in the limit of a large number of background sources, would be to 
select only RM distributions in the close proximity of dense substructures within filaments (which are presently masked out in our analysis), where small scale fluctuations in RMs are expected to be more significant compared to the fluctuations of the Galactic foreground and to the residual scatter of RMs.

Second,  we investigated a more optimistic primordial scenario, assuming a 10 times larger initial magnetic seed field (B$_{\rm init}$=1~nG) compared to our baseline primordial model. This case is meant to bracket the possibility that magnetic fields in real filaments are stronger than what is captured by MHD method, which may be potentially affected by resolution effects as any finite-volume method (see however discussion in \citet{va14mhd}). 
The  assumed B$_{\rm init}$=1~nG field is still below the most recent upper limit obtained by CMB analysis by \citet{PLANCK2015}. For simplicity we obtained this model by upscaling the magnetic field and the RM in our baseline model by $\times 10$ in post-processing, which is motivated by the fact that magnetic fields in filaments are expected not to be in the saturated dynamo regime \citep{ry08,2015MNRAS.453.3999M}.\\

\begin{figure}[H]
\centering
\includegraphics[width=0.49\textwidth]{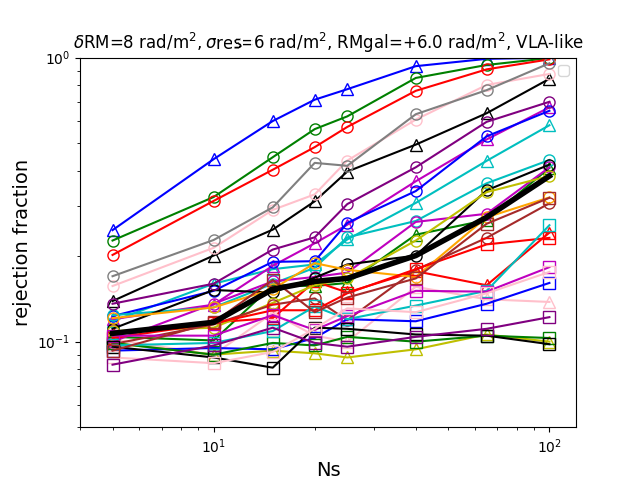}
\includegraphics[width=0.49\textwidth]{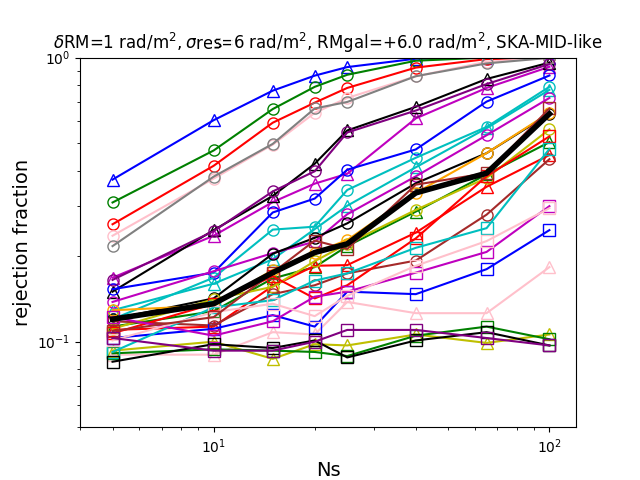}
\caption{Same as Fig.~\ref{fig:rej_frac} (lower panels) for the primordial seeding scenario with B$_{\rm init}=1$~nG.}
\label{fig:rej_frac_RMx10}
\end{figure}

For this optimistic case, the chances of detecting the magneto-ionic filamentary medium  are dramatically improved (see Fig.~\ref{fig:rej_frac_RMx10}). 
For example, setting $\sigma_{\rm res}=6$~rad/m$^2$ makes 
JVLA-like observations of filaments to reach a rejection fraction larger than 0.5 in $\sim 34\%$ of the sample with N$_S=100$ detected polarized sources, and $\sim 7\%$ with  N$_S=20$ sources.  With the SKA-MID more than a half of our objects will be significantly detectable with N$_S=100$ sources.  Even for a small number of available polarized sources N$_S=5$ about $\sim 10(17)\%$ of filaments have a chance of $\geq 1/5$ to be distinguished from the control field in a JVLA(SKA-MID)-like observation. \\

We conclude that filaments still represent a key case to test hypothesis on the initial magnetization level at large scales ($\geq$Mpc) since the magnetic field normalization plays a key role in their detection rate.

%%%%%%%%%%%%%%%%%%%%%%%%%%%%%%%%%%%%%%%%%%
\section{Conclusions}

%This section is not mandatory, but can be added to the manuscript if the discussion is unusually long or complex.

In this work, we assessed for the first time the possibility of detecting the cosmic web around massive filaments with radio polarisation observations and under realistic observing conditions. Our investigation has been motivated by the expected increase in performances of new large radio facilities.  
In particular, the planned sensitivity of the SKA will produce a leap in our knowledge about non-thermal constituents of the cosmic web. While continuum surveys should lead to the discovery of hundreds of new extended  radio sources in galaxy clusters  \citep[e.g.][]{2014arXiv1412.5940C,2017MNRAS.470..240N}, the expected flurry of accurate polarization data from extended and point-like sources should finally enable the detection of  the magnetized cosmic web around galaxy clusters, and to further study the topology and spatial distribution of magnetic fields on $\geq ~\rm Mpc$ scales, also representing a powerful new tool to locate and study the WHIM component of "missing baryons". \\

Unlike other works in the past, here we considered the real exercise that future radio observations will have to face, i.e. the challenge of statistically comparing distributions of RMs in putative filaments and in nearby control fields, by tacking into account realistic level of contamination from the Galaxy, different populations of background sources, and also including a realistic RM sensitivity. To assess the detection rate, we used the statistical M-W estimator \citep[][]{mw47}. \\

The results of our analysis show that pushing the instrumental RM sensitivity to $\leq \rm ~rad/m^2$ is key to probe magnetic fields in filaments, due to the higher statistics of detectable polarised  sources, which will allow us to distinguishing filaments from nearby control fields in a statistically robust way.
However, our analysis shows that strong limitations to this search arise because of the spurious contributions to the RM on the polarized radio sources \citep[e.g.][]{banfield14,2016ApJ...829....5L}.

These spurious contributions can be never fully avoided, but minimized  by selecting relatively nearby populations less affected by contributions to the RM along the LOS, or even populations showing smaller intrinsic RM scatter (e.g. WISE-Star radio sources \citep[also]{banfield14}) through cross-correlation with optical/infrared catalogs or self-consistently operating suitable cuts on their radio properties. 
In the scenario in which extragalactic fields have been mostly seeded by galaxies and active galactic nuclei, we expect an extremely small detection rate of filaments around galaxy clusters via RM studies.  Conversely, the systematic detection of RM from filaments (assuming the intrinsic contribution from sources can be opportunely minimized and the RM sensitivity is large enough) will provide an important  hints for significant levels of volume-filling magnetic fields already in the early Universe (as a result of a primordial magnetogenesis), or to an anomalously large amplification efficiency of large-scale motions in the WHIM. \\
Regardless of the magnetic seeding scenario, the filaments with the highest chances of detection via RM analysis in our sample are the ones with more gas substructures, which induce a higher level of compression and localized field amplification.\\

We caution that, since design the target of our analysis are filaments in rich environment populated by hundreds or thousands of galaxies, making exact prediction on the role of magnetic seeding by galaxies remains challenges as this depends on the still open issue of modeling star formation, AGN feedback and the interplay between galaxies and the magnetized intergalactic medium around them. 

However, our work strengthens also from an observational point of view the idea that the actual magnetization level of filaments (which in turn should depend on the magnetization history of the entire Universe) will crucially affect the success of future large polarization surveys in their attempt of detecting the Cosmic Web.

%%%%%%%%%%%%%%%%%%%%%%%%%%%%%%%%%%%%%%%%%%
\vspace{6pt} 

%%%%%%%%%%%%%%%%%%%%%%%%%%%%%%%%%%%%%%%%%%
%% optional
%\supplementary{The following are available online at www.mdpi.com/link, Figure S1: title, Table S1: title, Video S1: title.}

%%%%%%%%%%%%%%%%%%%%%%%%%%%%%%%%%%%%%%%%%%
\acknowledgments{In this work we used the {\enzo} code (http://enzo-project.org), the product of a collaborative effort of scientists at many universities and national laboratories.  Our simulations were run on the JUWELS cluster at Juelich Superc omputing Centre (JSC), under project account HHH42. We  acknowledge financial support from the Horizon 2020 program under the ERC Starting Grant "MAGCOW", no. 714196. We gratefully acknowledge B. Gaensler, A. Bonafede. M. Br\"{u}ggen, G. Sabatini and C. Gheller for fruitful scientific discussions and feedbacks.}

%%%%%%%%%%%%%%%%%%%%%%%%%%%%%%%%%%%%%%%%%%
\authorcontributions{N.L. and F.V. conceived and analysed the simulations; N.L. performed the statistical analysis; P.D.F. analyzed spectral distribution of magnetic energy; N.L. and F.V. wrote the paper.}

%%%%%%%%%%%%%%%%%%%%%%%%%%%%%%%%%%%%%%%%%%
\conflictofinterests{The authors declare no conflict of interest.} 

%%%%%%%%%%%%%%%%%%%%%%%%%%%%%%%%%%%%%%%%%%
%% optional
\abbreviations{The following abbreviations are used in this manuscript:\\

\noindent RM: Rotation Measure\\
SKA: Square Kilometer Array\\
MeerKAT: Karoo Array Telescope\\
JVLA: Jansky Very Large Array\\
K-S: Kolmogorov Smirnov\\  
M-W: Mann Whitney\\
$\Lambda$CDM: Lambda Cold Dark Matter\\
AGN: Active Galactic Nucleus\\
CMB: Cosmic Microwave Background\\
AMR: Adaptive Mesh Refinement\\
MHD: Magneto Hydro Dynamics\\
ICM: Intra Cluster Medium\\
IGM: Inter Galactic Medium\\
WHIM: Warm Hot Interagalactic Medium\\
LOS: Line of Sight}
%%%%%%%%%%%%%%%%%%%%%%%%%%%%%%%%%%%%%%%%%%
%% optional
\appendix
%\section{}
%The appendix is an optional section that can contain details and data supplemental to the main text. For example, explanations of experimental details that would disrupt the flow of the main text, but nonetheless remain crucial to understanding and reproducing the research shown; figures of replicates for experiments of which representative data is shown in the main text can be added here if brief, or as Supplementary data. Mathemtaical proofs of results not central to the paper can be added as an appendix.

%\section{}
%All appendix sections must be cited in the main text. In the appendixes, Figures, Tables, etc. should be labeled starting with `A', e.g., Figure A1, Figure A2, etc. 

%%%%%%%%%%%%%%%%%%%%%%%%%%%%%%%%%%%%%%%%%%
\bibliographystyle{apj}

%=====================================
% References, variant A: internal bibliography
%=====================================
%\renewcommand\bibname{References}
%\begin{thebibliography}{999}
% Reference 1
%\bibitem{ref-journal}
%Lastname, F.; Author, T. The title of the cited article. {\em %Journal Abbreviation} {\bf 2008}, {\em 10}, 142-149.
% Reference 2
%\bibitem{ref-book}
%Lastname, F.F.; Author, T. The title of the cited contribution. In {\em The Book Title}; Editor, F., Meditor, A., Eds.; Publishing House: City, Country, 2007; pp. 32-58.
%\end{thebibliography}

%=====================================
% References, variant B: external bibliography
%=====================================
\bibliography{franco2}

%%%%%%%%%%%%%%%%%%%%%%%%%%%%%%%%%%%%%%%%%%

%%%%%%%%%%%%%%%%%%%%%%%%%%%%%%%%%%%%%%%%%%
\end{document}